# Radiative effects of daily cloud cycle: general methodology and application to cloud fraction

Jun Yin[1,2]*, Amilcare Porporato[1,2]

1. Department of Civil and Environmental Engineering, Princeton University, Princeton, New Jersey, USA.

2. Princeton Environmental Institute, Princeton University, Princeton, New Jersey, USA.

*Correspondence to: jy12@princeton.edu, 609-258-5171

**OCRID**

Jun Yin: 0000-0003-2706-0620; Amilcare Porporato: 0000-0001-9378-207X

**Acknowledgments**

We acknowledge support from the USDA Agricultural Research Service cooperative agreement 58-6408-3-027; and National Science Foundation (NSF) grants EAR-1331846, EAR-1316258, FESD EAR-1338694 and the Duke WISeNet Grant DGE-1068871. The climate model data were downloaded from the fifth phase of the Coupled Model Intercomparison Project website (http://cmip-pcmdi.llnl.gov). The satellite observations from Clouds and the Earth's Radiant Energy System (CERES) were obtained from website (https://ceres.larc.nasa.gov/order_data.php). Models and codes used in the paper are available upon request.

**Abstract:**

The daily cloud cycle (DCC) and its response to global warming are critical to the Earth's energy budget, but their radiative effects have not been systematically quantified. Toward this goal, here we analyze the radiation at the top of the atmosphere and propose a measure of the DCC radiative effect (DCCRE) as the difference between the total radiative fluxes with the full cloud cycle and its uniformly distributed cloud counterpart. We apply it to the cloud fraction from four climate models that participated in the second phase of the Cloud Feedback Model Inter-comparison Project. The results allow us to objectively compare inter-model differences in the daily cycle of cloud fraction and their influences on the global energy balance. In particular, the spatial patterns of the DCC response to global warming obtained in our analysis suggest possible impacts on large-scale circulation. Such a framework can be used for conducting a more systematic evaluation of the DCC in climate models and observations with the goal to reduce uncertainty in climate projections.

## 1. Introduction

Cloud dynamics and their role in climate change are among the most important research areas in climate science (Stephens 2005; Bony et al. 2006; Zhang et al. 2013) and remain as one of the primary causes of uncertainty in climate projections (Cess et al. 1989; Boucher et al. 2013). Solving this difficult problem will allow to obtain more accurate simulations of cloud properties in the right place at the right time. While the spatial distribution of seasonal mean cloud properties has been extensively studied (Bony et al. 2006; Boucher et al. 2013), the behavior of cloud dynamics at sub-daily timescale, referred to as the daily cloud cycle (DCC), has attracted less attention (see however, Minnis and Harrison 1984; Bergman and Salby 1997; Yang and Slingo 2001; Clark et al. 2007; Taylor 2012; Pfeifroth et al. 2012; Langhans et al. 2013; Walther et al. 2013; Gustafson et al. 2014; Webb et al. 2015; Yin and Porporato 2017).

While it is rather obvious that an overcast sky may either tend to warm up the surface if clouds take place during the night or cool down the surface if clouds are during the day, the effects of subtle modulations in DCC on the radiative budget are difficult to assess. Multiple methods have been used to quantify the radiative effects of a climate variable and its response to an external perturbation. The cloud radiative effect (CRE) method (Cess et al. 1989, 1990, 1996) analyzes the effects of clouds by comparing all-sky and clear-sky radiative fluxes at the top of the atmosphere (TOA). This method is relatively easy to implement in global climate modes (GCMs), but the effects of the cloud properties are lumped together and confounded by other climate variables (Zhang et al. 1994; Colman 2003; Soden et al. 2004). The partial radiative perturbation (PRP) method (Wetherald and Manabe 1988) uses offline radiative transfer models to calculate the derivatives of TOA radiative fluxes with respect to a specified climate variable.

To facilitate its implementation, pre-calculated values of such derivatives, referred to as standard radiative kernels (Shell et al. 2008; Soden et al. 2008), have been used for estimating climate feedbacks without rerunning their own radiative transfer models (e.g. Vial et al. 2013). More recently, cloud radiative kernels based on the histograms of cloud fraction have been proposed to analyze the feedbacks of specific types of clouds (Zelinka et al. 2012a, b; Zhou et al. 2013). All these methods have been extremely useful in clarifying the inter-model variations of climate feedbacks; however, they have not been used for analyzing the radiative effects related to the DCC.

In this study, we draw from these methods and extend them to analyze the radiative effects of DCC. We consider the TOA radiative fluxes and isolate the DCC radiative effects from the effects of the daily mean cloud properties. We apply this method to the cloud-fraction results of four GCMs involved in the second phase of the Cloud Feedback Model Intercomparison Project (CFMIP-2). The paper is organized as follows: section 2 introduces the DCC and gives an example of daily cycle of cloud fraction. Section 3 analyzes its radiative effects by Taylor expanding the TOA radiative fluxes. A case study is presented in section 4 to analyze the daily cycle of cloud fraction and its response to global warming. Final conclusions are drawn in section 5.

## 2. Daily Cloud Cycle (DCC)

To analyze the DCC and its contribution to the Earth's energy budget, we first define the DCC as the long-term climatology of daily cloud cycle with reference to a generic cloud property, $c(t)$. We calculate $c(t)$ at each local time of day (e.g. 0, 1, …, 24 hour) for each season or month at a specific location, where $t$ is in local solar time within a day ($0 < t < 24$ hours). The cloud variable $c(t)$ is then decomposed into its mean, $c_0$, and fluctuations around it, $c_{\mathrm{DCC}}$. The latter is then decomposed into Fourier series as (Wood et al. 2002; Tian et al. 2006)

$$c(t) = c_0 + c_{\mathrm{DCC}}(t) = c_0 + \sum_{n=1}^{\infty} c_n \cos\left[nw(t - \phi_{c,n})\right], \qquad (1)$$

where, $c_n$ and $\phi_{c,n}$ are the amplitude and phase of the $n$th harmonic of the daily cycle, and $w = 2\pi/\tau$ is the angular frequency, in which $\tau$ is the length of one period of the daily cycle (i.e., 24 hours).

Figure 1 illustrates an example using for $c$ the cloud fraction ($f$) over the Atlantic Ocean near Puerto Rico as simulated by the GFDL-CM3. By comparing the current (1986-2005) and future (2081-2100) summertime cloud fraction climatology, the daily mean cloud fraction is seen to decrease ($\Delta f_0 = -2.52\%$), while the phase of the first harmonic shifts earlier ($\Delta \phi_{f,1} = -2.01$ hours) and the amplitude of the first harmonic increases ($\Delta f_1 = 1.03\%$).

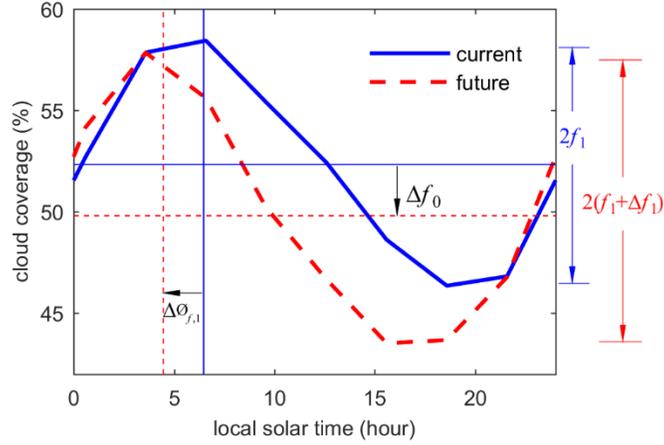

Figure 1. Example of daily cycle of cloud fraction near Puerto Rico over the North Atlantic Ocean (23N, 58.75W) in summer from GFDL-CM3 1986-2005 'historical' experiment (current) and 2081-2100 RCP45 experiment (future). The horizontal lines mark the mean cloud fraction $f_0$; the vertical lines refer to the phase of the first harmonic $\phi_{f,1}$; the heights of the dimension lines on the right of the figure indicate the change of twice the DCC amplitude of the first harmonic $f_1$.

## 3. DCC Radiative Effects

At a specific location and time of the year, the top-of-the-atmosphere (TOA) radiative flux $R$ in general depends on time of day $t$, as well as $c$ itself and other climate variables, say $x$. This TOA radiative flux $R(c,x,t)$ can be split into a contribution due to the mean and due to the DCC components. The former equals $R(c_0,x,t)$, while the remaining part is the DCC radiative effect:

$$\text{DCCRE}_c = R(c,x,t) - R(c_0,x,t). \tag{2}$$

We are interested in evaluating potential changes in DCCRE$_c$ corresponding to situations like the one shown in Figure 1. Thus, assuming a period of climatic change, the radiative impact of DCC response to such a perturbation can be written as

$$\Delta\text{DCCRE}_c = \Delta R_c - \Delta R_{c_0}, \tag{3}$$

where $\Delta R_c = R(c+\Delta c, x, t) - R(c, x, t)$ and $\Delta R_{c_0} = R(c_0 + \Delta c_0, x, t) - R(c_0, x, t)$. By expressing $c$ as its Fourier representation and Taylor expanding the TOA radiative flux, one obtains

$$\underbrace{R(c+\Delta c,x,t) - R(c,x,t)}_{\Delta R_c} = \underbrace{\frac{\partial R}{\partial c_0}\Delta c_0}_{\Delta R_{c_0}} + \sum_{n=1}^{\infty}\underbrace{\frac{\partial R}{\partial c_n}\Delta c_n}_{\Delta R_{c_n}} + \sum_{n=1}^{\infty}\underbrace{\frac{\partial R}{\partial \phi_{c,n}}\Delta\phi_{c,n}}_{\Delta R_{\phi_{c,n}}} + \text{H.O.T.}, \tag{4}$$

where H.O.T. are the higher order Taylor series terms due to the nonlinearity of the function $R(c,x,t)$. The first term on the r.h.s. of (4) is the radiative impact due to changes in mean cloud property ($\Delta R_{c_0}$), the second term refers to the impact from the amplitude modulation ($\Delta R_{c_n}$), and the third term represents the impact from phase shift ($\Delta R_{\phi_{c,n}}$). Rearranging equation (4) yields

$$\sum_{n=1}^{\infty} \Delta R_{c_n} + \sum_{n=1}^{\infty} \Delta R_{\phi_n} + \text{H.O.T.} = \Delta R_c - \Delta R_{c_0} = \Delta \text{DCCRE}_c, \quad (5)$$

which provides an interpretation of $\Delta \text{DCCRE}_c$ as the total radiative impacts of all the amplitude modulation, all the phase shift, and the corresponding higher order terms. Equations (2) and (5) closely resemble the formula used in CRE (Cess et al. 1989, 1990, 1996), which defines CRE as the all-day and clear-day radiative flux difference and estimates the total cloud radiative impacts as the change of CRE with the adjustment for cloud masking (Shell et al. 2008; Soden et al. 2008). Here instead, DCCRE$_c$ are first calculated considering the differences between radiative fluxes from the full-day cloud cycle and the fluxes from its counterpart assuming a uniformly redistributed $c$ equal $c_0$, and then computing the DCC radiative impacts for $c$ as the change of DCCRE$_c$.

Since $R(c_0, x, t)$ in Eq. (2) is usually not available from climate model outputs or from observations, to calculate DCCRE$_c$ using (5) one would need to rerun the radiative transfer codes in climate models. As an alternative to this direct calculation, one can approximate DCCRE$_c$ by taking the first term of the Taylor expansion around $c_0$

$$\text{DCCRE}_c = R(c,x,t) - R(c_0,x,t) \approx \frac{\partial R}{\partial c}(c - c_0), \quad (6)$$

where the derivative $\partial R / \partial c$ is the radiative kernel for $c$. In this way, the standard radiative kernels used to assess climate feedbacks (Shell et al. 2008; Soden et al. 2008; Zelinka et al. 2012a, b; Vial et al. 2013; Zhou et al. 2013) can be calculated at sub-daily timescale and readily be used to estimate the DCCRE. In the appendix, we also discuss an evaluation of DCCRE using a simplified energy balance model; this may be useful to provide quick estimates as well as to offer a transparent analytical tool to test hypotheses and explore the role of phase shift and amplitude modulation in DCC.

## 4. Application to radiative effects of daily cloud fraction

The cloud fraction (*f*) is of the most important cloud quantities and is critical to the Earth's energy budget (Stephens 2005; Bony et al. 2006; Boucher et al. 2013). Focusing on *f* has also the advantage that $R(f_0, x, t)$ in Eq. (2) can be easily obtained from CRE (Ramanathan et al. 1989; Cess et al. 1989, 1990, 1996) as

$$\mathrm{CRE}(t) = R(f,x,t) - R_{\mathrm{clr}}(x,t) = f(t)[R_{\mathrm{cld}}(x,t) - R_{\mathrm{clr}}(x,t)], \qquad (7)$$

where $R_{\mathrm{clr}}$ is the clear-sky radiative flux, and $R_{\mathrm{cld}}$ is the overcast cloudy-sky radiative flux. Rearranging (7) yields the TOA radiative fluxes

$$R(f,x,t) = f(t)R_{\mathrm{cld}}(x,t) + [1 - f(t)]R_{\mathrm{clr}}(x,t), \qquad (8)$$

so that

$$R(f_0,x,t) = f_0 R_{\mathrm{cld}}(x,t) + (1 - f_0)R_{\mathrm{clr}}(x,t). \qquad (9)$$

Combining (2), (7), (8), and (9) yields

$$\mathrm{DCCRE}_f = R(f,x,t) - R(f_0,x,t) = \frac{f(t) - f_0}{f(t)} \mathrm{CRE}(t). \qquad (10)$$

Note that both CRE and *f* at sub-daily time scale are required to calculate DCCRE*f*.

We apply Eq. (10) to four GCMs participated in the second phase of the Cloud Feedback Model Intercomparison Project (CFMIP-2) with half-hour (or three-hour) outputs at approximately 120 'cfSites' around the globe. These sites have large inter-model variability in cloud feedbacks and have been used for improving our understanding of cloud dynamics in GCMs (Bony et al. 2011; Taylor et al. 2011; Webb et al. 2015). We analyze the outputs from AMIP experiment forced with observed sea surface temperature and AMIP4K experiment with 4K sea surface temperature perturbation. Note that we focus on shortwave components because the DCC has much weaker impacts on the TOA longwave radiative fluxes (see Appendix and the weak daily cycle of longwave CRE in Fig. 3ab in Yin and Porporato 2017).

As in Figure 1, we calculate the daily cycle climatology of total cloud fraction from each 'cfSite' in each month from 1979-2010 for both AMIP and AMIP4K experiments. The obtained daily cycle climatology is then decomposed into Fourier series to find the mean, phase, and amplitude of the DCC. As shown in Figure 2 (a and b), the phase of the first harmonic over the ocean sites from all models are quite consistent but the phases over the continental sites from CNRM-CM5 are much closer to the noon. After 4K surface temperature perturbation in the AMIP4K experiment, the phase does not show significant change in all GCMs (see Figure 2 c and d). To assess how the inter-model variability of DCC influences the TOA radiative fluxes, we apply Eq. (10) to the daily cycle climatology of cloud fraction to calculate the DCCRE*f* for all 'cfSites' in each month. Figure 2 e and f show the shortwave $\mathrm{DCCRE}_f$ from the AMIP experiment averaged over all 12 months. The results show that clouds peaking at noon over the land in CNRM-CM5 (see Figure 2a) tend to reflect more solar radiation so that the corresponding DCCRE*f* have much lower values than those from other GCMs (see Figure 2e).

To investigate the DCC response to warming, we then calculated the change of its radiative effects ($\Delta\mathrm{DCCRE}_f$) from the AMIP experiment forced with observed sea surface temperature to the AMIP4K experiment with 4K perturbation. As shown in Figure 2 g and h, $\left|\Delta\mathrm{DCCRE}_f\right|$ are an order of magnitude smaller than $|\mathrm{DCCRE}_f|$. These results are also consistent with the study of Webb et al. (2015), which concludes that the daily cloud cycle may have a limited response to global warming.

Although the overall DCC response is weak, it may be significant in certain regions. Figure 3 presents the geographical patterns of $r\left(=\Delta \mathrm{DCCRE}_f / \Delta \mathrm{CRE}\right)$ in the month of February. Some locations have strong positive DCC radiative impacts, while there are strongly negative impacts in other places. Such spatial patterns, although resulting in a limited impact on the global mean energy budget, have the potential to modify the regional radiation balance and in turn influence large-scale circulations.

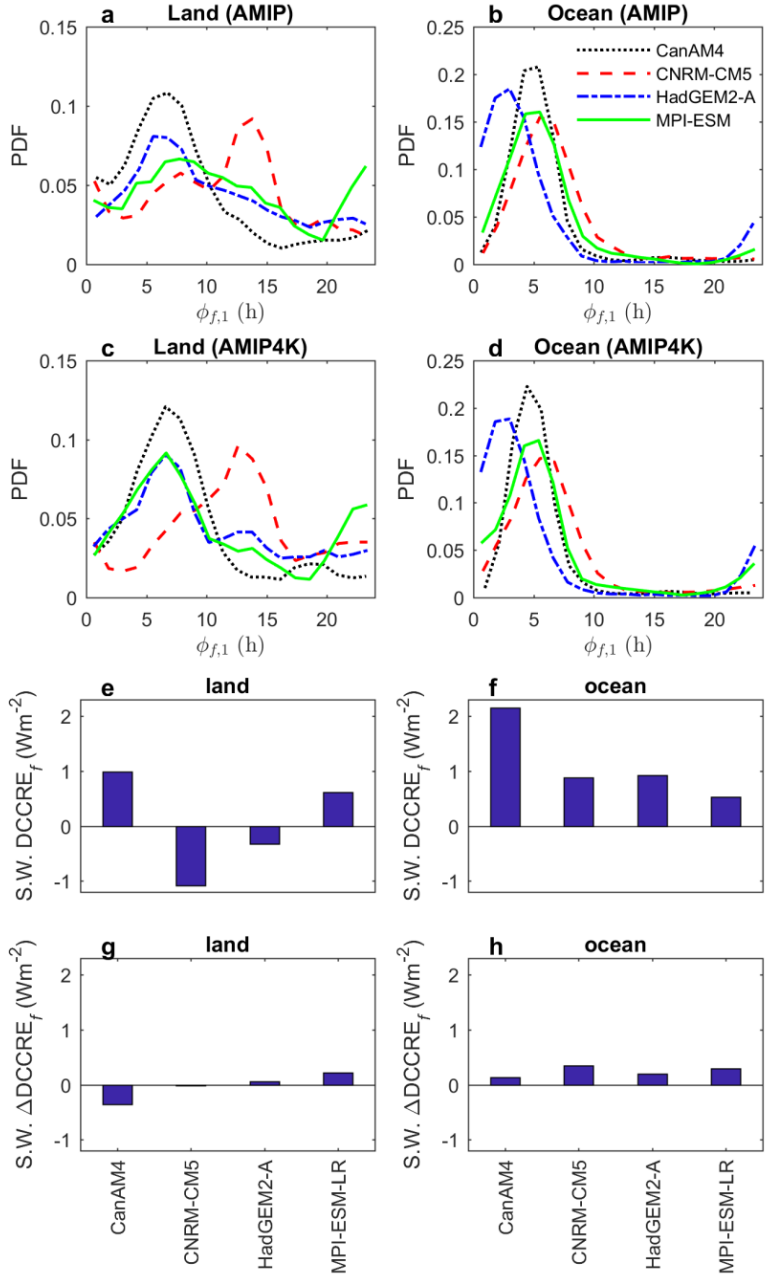

Figure 2. Daily cycle of cloud fraction and its radiative effects. The top four panels show the empirical probability density function (PDF) of the phase of the first harmonic of the cloud fraction, $\phi_{f,1}$, from (a and b) AMIP and (c and d) AMIP4K experiments over the (a and c) land and (b and d) ocean at all 'cfSites' in four GCMs. The bottom four panels are the corresponding shortwave $DCCRE_f$ and $\Delta DCCRE_f$. The $DCCRE_f$ have been averaged over all months over the (e) land and (f) ocean at the 'cfSites' during 1979-2010 from AMIP experiment; the $\Delta DCCRE_f$ over the (g) land and (h) ocean at 'cfSites' compares the $DCCRE_f$ from AMIP and AMIP4K experiments.

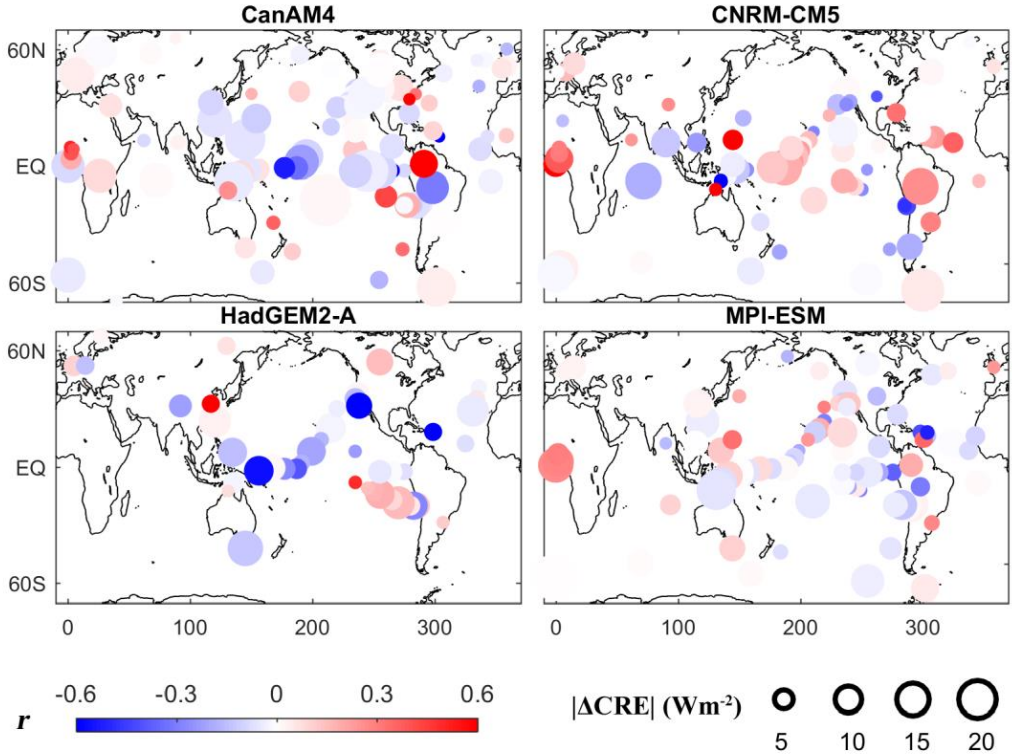

Figure 3. Spatial distribution of $r\,(=\Delta\text{DCCRE}/\Delta\text{CRE})$ from four GCM outputs at 'cfSites'. Location with $|\Delta\text{CRE}| < 2$ W m$^{-2}$ are removed which may have meaningless large $r$. The colors show the value of $r$ and the sizes represent the absolute values of $\Delta\text{CRE}$.

## 5. Conclusion

We have analyzed the daily cloud cycle radiative effects (DCCRE), expressed as $R(c,x,t) - R(c_0,x,t)$, which measures the differences of TOA radiative flux from full-day cloud cycle and from its counterpart with uniformly redistributed $c_0$. The proposed DCCRE method not only can be used to estimate the radiative impacts of DCC on current climate conditions but also can be applied to assess the DCC feedbacks. Our application to climate model outputs revealed large inter-model variations of radiative impacts of daily cycle of cloud fraction. The spatial patterns of DCC response to global warming in particular suggests a potential impacts on large-scale circulation, one of the key factors for climate projection (Stevens and Bony 2013; Stocker 2014).

The framework proposed is general to any cloud property. A comprehensive assessment of DCCRE could then be conducted once different cloud properties and their radiative kernels are available at sub-daily timescales. Such an assessment may provide valuable information to target particular aspects of cloud parametrization and the patterns of large-scale circulation, hopefully contributing to a better understanding of the uncertainties in climate projections.

**Appendix: Minimalist radiative balance model**

To provide a direct and relatively simple alternative to explore the radiative effects of TOA radiative fluxes, we also computed the radiative components using a minimalist radiative balance model (Lenton and Vaughan 2009; Kaper and Engler 2013; Hartmann 2015). Although simplified, the approach offers a transparent tool to test hypotheses and readily explore the role of phase shift and amplitude modulation in DCC.

For simplicity we consider a one-layer troposphere comprised of a fraction $f$ of clouds and a fraction 1-$f$ of greenhouse gas (see Figure A1).

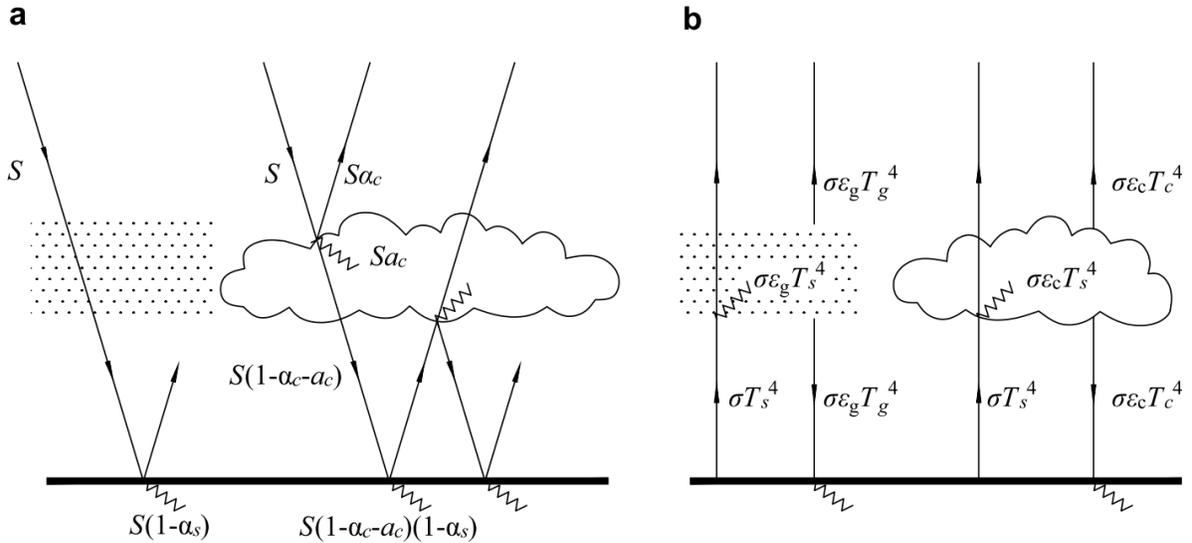

Figure A1 Schematic diagram of radiation components for the minimalist radiative balance model. The dotted area and cloud-shaped areas represent layers of clear and cloudy atmosphere, respectively. In (**a**), the solar radiative flux passes through the clear atmosphere, while it is partially reflected and absorbed by the clouds with albedo $\alpha_c$ and fractional absorption $a_c$. In (**b**), the clear atmosphere is assumed to be a grey body with emissivity $\varepsilon_g$, which absorbs and reemits the longwave radiation; the clouds also redistribute the longwave radiation in the same manner but with much higher emissivity $\varepsilon_c$.

The net shortwave radiation at the tropopause can be expressed as a combination of the shortwave radiations from cloudy/clear sky weighted by cloud fraction $f$

$$R_s(t) = [1 - f(t)]R_{s,\mathrm{clr}}(t) + f(t)R_{s,\mathrm{cld}}(t), \tag{A1}$$

where

$$R_{s,\text{clr}} = S(t)[1-\alpha_s], \tag{A2}$$

and

$$R_{s,\text{cld}} = S(t)(1-\alpha_c) - S(t)(1-\alpha_c-a_c)\alpha_s + \ldots = S(t)\left[(1-\alpha_c) - (1-\alpha_c-a_c)^2 \frac{\alpha_s}{1-\alpha_s\alpha_c}\right], \tag{A3}$$

where $\alpha_c$ and $a_c$ are cloud albedo and fractional absorption, $\alpha_s$ is surface albedo, and $S$ is the solar radiative flux reaching the tropopause.

The net longwave radiative flux is

$$R_l(t) = [1-f(t)]R_{l,\text{clr}}(t) - f(t)R_{l,\text{cld}}(t), \tag{A4}$$

where

$$R_{l,\text{clr}} = -\sigma(1-\varepsilon_g)T_s^4(t) - \sigma\varepsilon_g T_g^4(t), \tag{A5}$$

and

$$R_{l,\text{cld}} = -\sigma(1-\varepsilon_c)T_s^4(t) - \sigma\varepsilon_c T_c^4(t), \tag{A6}$$

where $\sigma$ is Stefan–Boltzmann constant, $\varepsilon_g$ is bulk longwave emissivity of the simplified greenhouse gas layer in the clear sky, $\varepsilon_c$ is the bulk longwave emissivity of the cloudy atmosphere, and $T_c$, $T_g$, and $T_s$ are the temperature of the cloudy layer, the greenhouse gas layer, and the earth surface.

The impact of the DCC on radiation can now be illustrated with reference to the example in Figure 1. The ocean surface albedo is set as 0.1, while the cloud albedo and shortwave emissivity are modeled using well-established empirical functions of cloud water path and solar zenith angle (Stephens 1978). An in-cloud water path of 113 g m$^{-2}$ and the daily cycle of sea surface temperature are obtained from the GFDL-CM3 outputs. For the longwave radiation, the energy in the atmospheric layer is assumed to be in equilibrium ($T_g = T_c = 1/2^{1/4}T_s$) (Bohren and Clothiaux 2006; Hartmann 2015). The clouds are assumed to behave like a blackbody due to the fact that water drops are efficient absorbers/emitters of longwave radiation, and the clear atmosphere is assumed to have emissivity of 0.78 (Bohren and Clothiaux 2006).

Figure A2 shows the impacts of phase and amplitude of the first harmonic on the daily-averaged radiative fluxes for the example in Figure 1. The radiative impacts of phase $\phi_{t,1}$ on both shortwave and longwave radiations are more significant under larger amplitude $f_1$. With constant daily mean value $f_0$, clouds peaking at noon reflect much more solar radiation, while longwave emission from surface peaks in the afternoon due to the maximum sea surface temperature. When clouds peak at this time, slightly more longwave radiation is trapped, as the daily surface temperature amplitude is only about 0.1K. The impacts of DCC on longwave radiation are negligible, unless the cloud structure has significant daily variations (Bergman and Salby 1997; Taylor 2012).

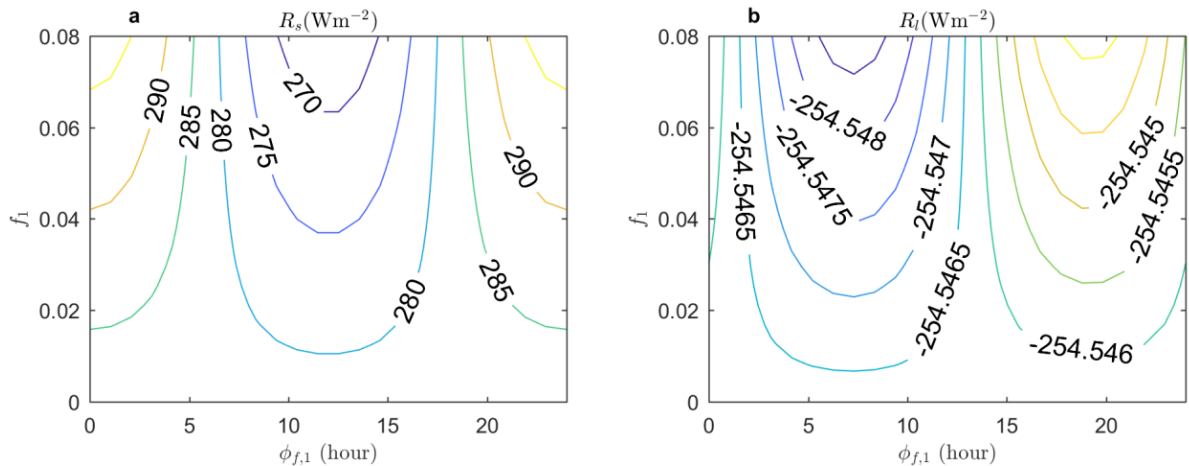

Figure A2 Daily-averaged TOA (a) shortwave and (b) longwave radiative fluxes as functions of amplitude and phase of the first harmonic of the cloud fraction for the example in Figure 1. The gradients show the impacts of DCC amplitude and phase has strong impact on the shortwave radiation but negligible influence on the longwave one.

**References**

Bergman JW, Salby ML (1997) The Role of Cloud Diurnal Variations in the Time-Mean Energy Budget. J Clim 10:1114–1124. doi: 10.1175/1520-0442(1997)010<1114:TROCDV>2.0.CO;2

Bohren CF, Clothiaux EE (2006) Fundamentals of atmospheric radiation: an introduction with 400 problems. John Wiley & Sons

Bony S, Colman R, Kattsov VM, et al (2006) How Well Do We Understand and Evaluate Climate Change Feedback Processes? J Clim 19:3445–3482. doi: 10.1175/JCLI3819.1

Bony S, Webb M, Bretherton CS, et al (2011) CFMIP: Towards a better evaluation and understanding of clouds and cloud feedbacks in CMIP5 models. Clivar Exch 56:20–22

Boucher O, Randall D, Artaxo P, et al (2013) Clouds and Aerosols. In: Stocker TF, Qin D, Plattner G-K, et al. (eds) Climate Change 2013: The Physical Science Basis. Contribution of Working Group I to the Fifth Assessment Report of the Intergovernmental Panel on Climate Change. Cambridge University Press, Cambridge, United Kingdom and New York, NY, USA, pp 571–658

Cess RD, Potter GL, Blanchet JP, et al (1989) Interpretation of Cloud-Climate Feedback as Produced by 14 Atmospheric General Circulation Models. Science 245:513–516. doi: 10.1126/science.245.4917.513

Cess RD, Potter GL, Blanchet JP, et al (1990) Intercomparison and interpretation of climate feedback processes in 19 atmospheric general circulation models. J Geophys Res Atmospheres 95:16601–16615. doi: 10.1029/JD095iD10p16601

Cess RD, Zhang MH, Ingram WJ, et al (1996) Cloud feedback in atmospheric general circulation models: An update. J Geophys Res Atmospheres 101:12791–12794. doi: 10.1029/96JD00822

Clark AJ, Gallus WA, Chen T-C (2007) Comparison of the Diurnal Precipitation Cycle in Convection-Resolving and Non-Convection-Resolving Mesoscale Models. Mon Weather Rev 135:3456–3473. doi: 10.1175/MWR3467.1

Colman R (2003) A comparison of climate feedbacks in general circulation models. Clim Dyn 20:865–873. doi: 10.1007/s00382-003-0310-z

Gustafson WI, Ma P-L, Singh B (2014) Precipitation characteristics of CAM5 physics at mesoscale resolution during MC3E and the impact of convective timescale choice. J Adv Model Earth Syst 6:1271–1287. doi: 10.1002/2014MS000334

Hartmann DL (2015) Global physical climatology. Newnes

Kaper H, Engler H (2013) Mathematics and climate. SIAM

Langhans W, Schmidli J, Fuhrer O, et al (2013) Long-Term Simulations of Thermally Driven Flows and Orographic Convection at Convection-Parameterizing and Cloud-Resolving Resolutions. J Appl Meteorol Climatol 52:1490–1510. doi: 10.1175/JAMC-D-12-0167.1

Lenton TM, Vaughan NE (2009) The radiative forcing potential of different climate geoengineering options. Atmospheric Chem Phys 9:5539–5561

Minnis P, Harrison EF (1984) Diurnal Variability of Regional Cloud and Clear-Sky Radiative Parameters Derived from GOES Data. Part III: November 1978 Radiative Parameters. J Clim Appl Meteorol 23:1032–1051. doi: 10.1175/1520-0450(1984)023<1032:DVORCA>2.0.CO;2

Pfeifroth U, Hollmann R, Ahrens B (2012) Cloud Cover Diurnal Cycles in Satellite Data and Regional Climate Model Simulations. Meteorol Z 21:551–560. doi: 10.1127/0941-2948/2012/0423

Ramanathan V, Cess RD, Harrison EF, et al (1989) Cloud-radiative forcing and climate: results from the Earth radiation budget experiment. Science 243:57–63. doi: 10.1126/science.243.4887.57

Shell KM, Kiehl JT, Shields CA (2008) Using the Radiative Kernel Technique to Calculate Climate Feedbacks in NCAR's Community Atmospheric Model. J Clim 21:2269–2282. doi: 10.1175/2007JCLI2044.1

Soden BJ, Broccoli AJ, Hemler RS (2004) On the Use of Cloud Forcing to Estimate Cloud Feedback. J Clim 17:3661–3665. doi: 10.1175/1520-0442(2004)017<3661:OTUOCF>2.0.CO;2

Soden BJ, Held IM, Colman R, et al (2008) Quantifying Climate Feedbacks Using Radiative Kernels. J Clim 21:3504–3520. doi: 10.1175/2007JCLI2110.1

Stephens GL (1978) Radiation Profiles in Extended Water Clouds. II: Parameterization Schemes. J Atmospheric Sci 35:2123–2132. doi: 10.1175/1520-0469(1978)035<2123:RPIEWC>2.0.CO;2

Stephens GL (2005) Cloud Feedbacks in the Climate System: A Critical Review. J Clim 18:237–273. doi: 10.1175/JCLI-3243.1

Stevens B, Bony S (2013) What Are Climate Models Missing? Science 340:1053–1054. doi: 10.1126/science.1237554

Stocker T (2014) Climate change 2013: the physical science basis: Working Group I contribution to the Fifth assessment report of the Intergovernmental Panel on Climate Change. Cambridge University Press

Taylor KE, Stouffer RJ, Meehl GA (2011) An Overview of CMIP5 and the Experiment Design. Bull Am Meteorol Soc 93:485–498. doi: 10.1175/BAMS-D-11-00094.1

Taylor PC (2012) Tropical outgoing longwave radiation and longwave cloud forcing diurnal cycles from CERES. J Atmospheric Sci 69:3652–3669

Tian B, Waliser DE, Fetzer EJ (2006) Modulation of the diurnal cycle of tropical deep convective clouds by the MJO. Geophys Res Lett 33:n/a-n/a. doi: 10.1029/2006GL027752

Vial J, Dufresne J-L, Bony S (2013) On the interpretation of inter-model spread in CMIP5 climate sensitivity estimates. Clim Dyn 41:3339–3362. doi: 10.1007/s00382-013-1725-9

Walther A, Jeong J-H, Nikulin G, et al (2013) Evaluation of the warm season diurnal cycle of precipitation over Sweden simulated by the Rossby Centre regional climate model RCA3. Atmospheric Res 119:131–139. doi: 10.1016/j.atmosres.2011.10.012

Webb MJ, Lock AP, Bodas-Salcedo A, et al (2015) The diurnal cycle of marine cloud feedback in climate models. Clim Dyn 44:1419–1436. doi: 10.1007/s00382-014-2234-1

Wetherald RT, Manabe S (1988) Cloud Feedback Processes in a General Circulation Model. J Atmospheric Sci 45:1397–1416. doi: 10.1175/1520-0469(1988)045<1397:CFPIAG>2.0.CO;2

Wood R, Bretherton CS, Hartmann DL (2002) Diurnal cycle of liquid water path over the subtropical and tropical oceans. Geophys Res Lett 29:7-1-7–4. doi: 10.1029/2002GL015371

Yang G-Y, Slingo J (2001) The Diurnal Cycle in the Tropics. Mon Weather Rev 129:784–801. doi: 10.1175/1520-0493(2001)129<0784:TDCITT>2.0.CO;2

Yin J, Porporato A (2017) Diurnal cloud cycle biases in climate models. Nat Commun 8:2269. doi: 10.1038/s41467-017-02369-4

Zelinka MD, Klein SA, Hartmann DL (2012a) Computing and Partitioning Cloud Feedbacks Using Cloud Property Histograms. Part I: Cloud Radiative Kernels. J Clim 25:3715–3735. doi: 10.1175/jcli-d-11-00248.1

Zelinka MD, Klein SA, Hartmann DL (2012b) Computing and Partitioning Cloud Feedbacks Using Cloud Property Histograms. Part II: Attribution to Changes in Cloud Amount, Altitude, and Optical Depth. J Clim 25:3736–3754. doi: 10.1175/jcli-d-11-00249.1

Zhang M, Bretherton CS, Blossey PN, et al (2013) CGILS: Results from the first phase of an international project to understand the physical mechanisms of low cloud feedbacks in single column models. J Adv Model Earth Syst 5:826–842. doi: 10.1002/2013MS000246

Zhang MH, Hack JJ, Kiehl JT, Cess RD (1994) Diagnostic study of climate feedback processes in atmospheric general circulation models. J Geophys Res Atmospheres 99:5525–5537. doi: 10.1029/93JD03523

Zhou C, Zelinka MD, Dessler AE, Yang P (2013) An Analysis of the Short-Term Cloud Feedback Using MODIS Data. J Clim 26:4803–4815. doi: 10.1175/JCLI-D-12-00547.1